\documentclass[twocolumn,showpacs,pra,aps]{revtex4}

\usepackage{footnote}
\usepackage{dcolumn}

\newcommand{\cd}{\!\cdot\! }

\begin{document}

\title{On Status of Boltzmann Kinetic Theory \\
in the Framework of Statistical Mechanics}

\author{Yuriy E. Kuzovlev}
\email{kuzovlev@fti.dn.ua} \affiliation{Donetsk Institute
for Physics and Technology NASU, 83114 Donetsk, Ukraine}


\begin{abstract}
It is shown that early suggested derivation of the Boltzmann kinetic
equation for dilute hard sphere gas from the time-reversible BBGKY
equations is incorrect since in fact a priori substitutes for them
definite irreversible equations. Alternative approach to analysis of
the hard sphere gas is formulated which conserves the reversibility
and makes it clear that at any gas density one can reduce the BBGKY
equations to the Boltzmann equation only in case of spatially uniform
gas.
\end{abstract}

\pacs{05.20.-y, 05.40.-a}

\maketitle

{\bf 1}.\, The Boltzmann kinetic equation (BE) \cite{bol,re,lp,ub} is
one of most beautiful and fruitful models of theoretical physics.
However, its status from viewpoint of statistical mechanics still
stays under question. By the conventional opinion, solutions of BE
coincide with solutions of exact equations of statistical mechanics,
i.e. Bogolyubov-Born-Green-Kirkwood-Yvon (BBGKY) equations (BBGKYE)
\cite{bog}, at least in the low-density gas limit or in the
mathematically equivalent Boltzmann-Grad limit (BGL) when $\,a
\rightarrow 0\,$, $\,\nu \rightarrow\infty\,$,\, $\,\mu =a^3\nu
\rightarrow 0\,$,\, $\, \lambda  =(\pi a^2\nu )^{-1}=\,$const\,, with
$\,\nu\,$ being mean gas density (concentration of gas particles) and
$\,a\,$ and $\,\lambda\,$ being interaction radius and mean free path
of gas particles, respectively \cite{ub,lan}. The attempt to prove
this assumption by considering the hard (elastic) sphere gas was made
by Lanford and is known as ``Lanford theorem'' \cite{lan,vblls,pg}.
But the Lanford result does not seem to be quite convincing because
it was based on the formal series of iterations of BBGKYE which
converges for absurdly small evolution times only, $\,t< \tau\,$
($\,\tau\sim \lambda/\sqrt{T/m}\,$ is mean free path time).
Nevertheless, the ``Lanford theorem'' hardly is compatible with
results of \cite{i1} (see also \cite{i2,p12,tmf,p3,igp}) where for
the gas of ``soft elastic spheres'' it was shown that in case of its
spatial non-uniformity BE does not follow from BBGKYE even under BGL.

The aim of the present paper is to reveal the origin of this
contradiction starting from the ``hard sphere BBGKY hierarchy''
 \cite{re,lan,pg,ch,hs}. We will see that such the method for building
solutions to this hierarchy as applied in \cite{lan,vblls,pg}
destroys its symmetry in respect to time inversion and insensibly
replaces it by definite kinetic, i.e. irreversible, equations.
Therefore the results of \cite{lan,vblls,pg} can not be qualified as
BE derivation from BBGKYE. Besides, we will suggest and discuss a new
approach to analysis of true solutions of the hard sphere BBGKY
hierarchy.

{\bf 2}.\, Let us consider the hard sphere gas \cite{pg,ch}. There
are no rigorous rules for transition to it from gas with smooth
inter-particle interaction (thought a non-rigorous procedure was
considered e.g. in \cite{hs}). But it is possible first to postulate
the Liouville equation for the hard sphere system, as combination of
Liouville equation for free particles,\,
\[
\begin{array}{c}
\partial F/\partial t =-\sum_i {\bf v}_i\cd \nabla_i F\,\,\,\,\,
\,\,\, \texttt{at} \,\,\,\,\, \,|{\bf r}_i-{\bf r}_j|
> a\,
\end{array}
\]
(where\, $\,{\bf v}_i  ={\bf p}_i/m\,$, $\,\nabla_i
=\partial/\partial {\bf r}_i\,$), and boundary conditions to it,
\begin{equation}
\begin{array}{c}
F(...{\bf p}_i^* ...{\bf p}_j^*...) = F(...{\bf p}_i ...{\bf
p}_j...)\,\, \,\, \texttt{at} \,\,\,\, {\bf r}_j-{\bf r}_i
=a\Omega\,\,,\\ {\bf p}^*_{i,j}\,=\, {\bf p}_{i,j}\,\pm \,\Omega
\,(\Omega\cd  ({\bf p}_j -{\bf p}_i)) \,\, \label{mr}
\end{array}
\end{equation}
($\,\Omega\,$ is unit vector),\, which establishe continuity of
(density of) probability measure $\,F\,$ along phase trajectories of
particles under their collisions. Then, second, from here one can in
usual way \cite{bog} deduce the desirable BBGKYE:
\begin{eqnarray}
\frac {\partial F_{n}}{\partial t}= - \sum_{j\,=1}^{n} {\bf v}_j\cd
\,\nabla_j\,\,F_n \,+\,\nu \sum_{j\,=1}^{n}
\widehat{\mathbf{I}}_{j,n+1}\,\, F_{n+1}\, \,\,, \,\,\,\, \,
\label{fn}
\end{eqnarray}
where satisfaction of the boundary conditions (\ref{mr}) is presumed,
and the ``collision operators'' are defined by
\begin{eqnarray}
\widehat{\mathbf{I}}_{j,k}  F =
 a^2\!\! \int \!\!\int \Omega\!\cd \!({\bf
v}_{k}\! -\! {\bf v}_j)\, F({\bf r}_{k}={\bf r}_j\! +a\Omega )\,
d{\bf p}_{k}\,d\Omega   \,\, \, \label{ci0}
\end{eqnarray}

It should be emphasized that these equations, like BBGKY in general,
are time-reversible:\, if $\,\{F_n(t,\,{\bf r},{\bf p})\}\,$ is some
solution to equations (\ref{mr})-(\ref{ci0}) then
$\,\{F_n(-\,t,\,{\bf r},-\,{\bf p})\}\,$ also is their solution.

At this point the serious question does arise: how we have to deal
with the conditions (\ref{mr})? For the first look, we can merely use
these conditions to express probabilities of post-collision ({\it
\,out}-) states via probabilities of pre-collision ({\it \,in}-)
states and after that exclude the conditions from consideration.

Then the collision operators take the form
\begin{eqnarray}
\,\,\, \widehat{\mathbf{I}}_{j,k}\, F  =\,a^2 \int \!\!\int
(-\Omega\!\cd \!({\bf v}_{k} -{\bf v}_j))\,\, \theta( - \Omega\!\cd
\!({\bf v}_{k} -{\bf v}_j)) \, \times
\nonumber  \\
\times\,\, [\, F({\bf r}_k={\bf r}_j-a\Omega\,, {\bf p}^*_j\,,{\bf
p}^*_k\, ) \,-\, \,\,\,\, \, \,\,\,\, \, \,\,\,\, \, \,\,\,\, \,\,\,
\label{bci} \\ \,  \,\,\,\, \,\, \, \,\,\,\, -\, F({\bf r}_k={\bf
r}_j+a\Omega\,, {\bf p}_j\,,{\bf p}_k\,) \,]\, \,d{\bf p}_k\,d\Omega
\,\,\,, \nonumber
\end{eqnarray}
where $\,\theta(\cd )\,$ is the Heavyside function indicating that
the integration includes {\it \,in}-states only.

Just such transformed equations conventionally are assumed as a basis
of the theory. All the more, they a priori are well predisposed to
the Boltzmann's Sto{\ss}zahlansatz \cite{bol}. Indeed, if we truncate
the transformed hierarchy of equations at $\,n=s\,$, neglecting
$\,(s+1)$-particle correlations, then the residuary $\,s\,$
equations, - e.g.
\begin{eqnarray}
\frac {\partial F_1}{\partial t}= - {\bf v}_1\! \cd \! \nabla_1\,F_1
+\nu  a^2\!\! \int\!\! d{\bf p}_2\!\!\int \! d\Omega\, (\Omega\!\cd
\!{\bf v}_{21}) \, \theta(\Omega\! \cd \!{\bf v}_{21})
 \times \nonumber \\
 \times  [ F_2({\bf r}_{21}\!=-a\Omega, {\bf p}^*_1, {\bf
p}_2^*) - F_2({\bf r}_{21}\!=a\Omega,{\bf p}_1,{\bf p}_2)] \,\,,
\nonumber \\
\frac {\partial F_2}{\partial t}= - \!\!\sum_{j=1}^{2} {\bf v}_j\!\cd
\! \nabla_j\,F_2+\sum_{j=1}^{2} \nu a^2 \int\!\! d{\bf p}_3\!\!\int
\! d\Omega\, (\Omega\!\cd \!{\bf v}_{3j}) \times\,\,
\nonumber\\
\times\,\theta(\Omega\! \cd \!{\bf v}_{3j})\,
 [ F_2({\bf p}^*_j)\,F_1({\bf r}_{3}\!={\bf r}_{21}-a\Omega,
{\bf p}^*_3)\,- \,\,\,\,\,\,\,\,\,\label{f2}
\\
\,\,\,\, \,\,\,\,-\,  F_2({\bf p}_j)\,F_1({\bf r}_{3}\!={\bf
r}_{21}+a\Omega, {\bf p}_3)]\,\,\,\,  \nonumber
\end{eqnarray}
at $\,s=2\,$\,\footnote{\, At $\,s=1\,$ that is the Boltzmann-Enskog
equation \cite{re,bnn}.}\,, - under the BGL directly lead to the
BE\,\footnote{\, In more detail, if at $\,t=0\,$ the ``mplecular
chaos'' takes place, $\,F_n(t=0,{\bf r},{\bf p}) =\prod_{j=1}^n
F_1^0({\bf r}_j,{\bf p}_j)\,$, then under BGL it propagates to
 $\,t>0\,$, with $\,F_1(t)\,$ satisfying BE.}\,,
\begin{eqnarray}
\frac {\partial F_1}{\partial t}= - {\bf v}_1\! \cd \! \nabla_1\,F_1
+\nu  a^2\! \int\! \!\int  (\Omega\!\cd \!{\bf v}_{21}) \,
\theta(\Omega\! \cd \!{\bf v}_{21})
\, \times \,\,\,\, \,\label{ub} \\
\,\, \times \, [\, F_1({\bf p}^*_1)\, F_1({\bf p}_2^*) - F_1({\bf
p}_1)\, F_1({\bf p}_2)\,] \, d\Omega\,d{\bf p}_2\, \,\, \nonumber
\end{eqnarray}
(we introduced\, $\,{\bf v}_{ij}={\bf v}_{i} -{\bf v}_j\,$).\, This
statement is in essence equal to the `Lanford theorem''\,\footnote{\,
Formally, it is even stronger than the Lanford theorem since is not
restricted in respect to the evolution time.}\,, because behavior of
any term of the iteration series \cite{lan,pg} under BGL is
determined by a finite number of the transformed equations.

{\bf 3}.\, Notice, however, that the transformation of Eqs.\ref{fn},
by substituting (\ref{bci}) for (\ref{ci0}), takes away from them
their aforesaid time-reversibility property: now they are not
invariant in respect to the inversion\, $\,t\rightarrow -t\,,\, {\bf
p}\rightarrow -{\bf p}\,$. Consequently solutions of the transformed
equations in general are not solutions of the initial BBGKYE, and
vise versa. In other words, that non-equivalent hierarchies of
equations!

Or, to be more precise, probably, we would be able to consider them
as equivalent if we were able to deal simultaneously with all
possible many-particle correlations (of arbitrary high orders). But
they are certainly non-equivalent if we neglect even a part of
correlations\,\footnote{\, For example, irreversibility of equations
(\ref{f2}) is obvious.}\,.

Therefore the term by term consideration of iteration series for the
transformed equations \cite{lan,pg}  leads away from a true solution
of the initial ``hard sphere BBGKY hierarchy''
(\ref{mr})-(\ref{ci0}). Instead it offers solution of another, in
essence, postulated kinetic equations. Seemingly, this result sorts
with true BBGKY solutions approximately like complex-energy solutions
of the Schrodinger equations \cite{bzp} (or, in classical
statmechanics, solutions of the Liouville  equations corresponding to
the complex ``Ruelle-Pollicott resonances'' \cite{gasp}) sort with
their actual solutions\,\footnote{\, All these resemble also the
``strange'' phenomenon of ``unitary non-equivalence'' of different
mathematical descriptions of one and the same physical situation
\cite{emh}.}\,.

{\bf 4}.\, Thus, the replacement of (\ref{ci0}) by (\ref{bci}) gives
the same effect as the Boltzmann's Sto{\ss}zahlansatz. From the
physical point of view, this ``ansatz'' is bad as for it forces us to
neglect pre-collision ({\it \,in\,}-) inter-particle correlations and
hence neglect fluctuations in ``relative frequency'', or ``frequency
ratio'' (``chastost''' in Russian \cite{kr}) of collisions
\cite{i1,i2}.

Indeed, any collision starts from a pre-collision configuration and
finishes with a post-collision configuration, hence, fluctuations in
``relative frequency of collisions'' equally give rise to both {\it
\,out\,}-\, and {\it \,in\,}-correlations\,\footnote{\, In addition,
since pre-collision configurations arise by time $\,\sim\tau \,$
earlier than really collision at distance  $\,\sim\lambda\,$ from it
and post-collision configurations disappear at time $\,\sim\tau \,$
after it at distance  $\,\sim\lambda\,$ from it, then both {\it
\,out\,}-\, and  {\it \,in\,}-correlations enclose volumes $\,\sim
\pi a^2\lambda =1/\nu\,$ \cite{tmf,p3,igp}.}\,.

It is necessary to emphasize that we say about statistical
correlations which do not presume presence of some
cause-and-consequence relations beyond them\,\footnote{\, Mutual
independence of colliding particles in the sense of absence of some
mutual prehistory generally does not mean statistical independence of
the particles in the sense of the probability theory. \newline Citing
\cite{kr}, ``{\it That are prejudices ... that phenomena which are
``obviously independent'' should possess independent probability
distribution laws}\,'', or\, ``{\it that for any phenomenon one
always can point out some definite probability value}\,'' (this is my
own translation from Krylov's original Russian text). \newline And
more: ``\,{\it ... relative frequencies of one or another phenomenon
along phase trajectory, generally speaking, in no way are related to
their a priori probabilities}\,''.}\,.

The cause of the ``fluctuations in relative frequency of collisions''
is mere absence of back reaction to them when they do not disturb the
system's state (for instance, when relative frequencies of mutually
time-reversed collisions fluctuate with keeping definite proportions
between them)\,\footnote{\, Such kind of fluctuations for the first
timr were discussed in \cite{bk12,bk3}, later in application to
fluids in \cite{i1,p12,i2} and in application to other systems in in
\cite{i2,i3,i4}, and besides recently in \cite{tmf,p3,igp}. In
\cite{i1} it was shown that  these fluctuations are indifferent to a
degree of smallness of the gas parameter  $\,\mu\,$. In opposite,
just at\, $\,\mu\rightarrow 0\,$ (in BGL) the ``absence of back
reaction'' is especially easy understandable. \newline There is a key
to understanding the 1/f-noise observed in various physical systems
\cite{i1,p12,i2,bk12,bk3,i3,i4}. And we can expect that exact
time-reversible solutions of BBGKYE contain 1/f fluctuations in
kinetic characteristics of the system.}\,.

Clearly, these fluctuations are as well reflected by the distribution
functions (DF) $\,\{F_n(t)\}\,$ as strong is spatial non-uniformity
of the system \cite{i1,i2}, and therefore they are reflected in the
form of correlations between particles' coordinates (while their
velocities can be uncorrelated as in the Boltzmann's
theory)\,\footnote{\, By this reason, a correct derivation of BE from
BBGKYE is possible only for uniform gas. This was claimed in
\cite{kac} on those ground that in non-uniform case the averaging
over statistical ensemble can not be replaced by averaging over gas
volume. }\,.

These spatial correlations\,\footnote{\, Or, better saying,
configurational correlations, since in general they are dependent on
small details of relative particle's dispositions at scales
$\,\lesssim a\,$.}\,, in their turn, do mean that the DF values  at
collision configurations, e.g. $\,F_2({\bf r}_2={\bf r}_1+a\Omega\,$,
represent independent on  $\,F_1(t)\,$ and complementary to
$\,F_1(t)\,$ characteristics of statistical ensemble\,\footnote{\,
Already because for ensemble averages generally the inequalities \,
$\,\langle \widetilde{\nu}^s\rangle \neq\langle
\widetilde{\nu}\rangle^s\,$\,  take place, where
$\,\widetilde{\nu}\,$ is local gas density.}\,.

{\bf 5}.\, In view of the aforesaid, we have to come back to the
question how we must deal with the conditions (\ref{mr}).

Since ``contact'' DF's values which enter (\ref{mr}), first of all
$\,F_2({\bf r}_1 ={\bf r}_2+a\Omega\,$, play the role of ``governing
parameters'' for BBGKY hierarchy (\ref{fn}) as the whole, it is
natural to treat them as independent on $\,F_1\,$ characteristics of
gas. In more detail, when considering $\,F_2({\bf r}_1 ={\bf
r}_2+a\Omega\,$, we inevitably come to rest (in the collision
integral) against three-particle configurations corresponding to
pairs of infinitely close [air collisions. The, considering such
configurations, we will come to analogous four-particle ones, and so
on. Categorizing all them, one would construct a full (infinite)
system of equations for the ``contact'' Dfs. Such a system, of
course, would be time-reversible\,\footnote{\, Just by this reason it
would present true statistical weights of any of kinematically
possible scenarios of collisions.  \newline Generally, since
irreversibility equally manifests itself in both opposite time
directions, its completely adequate description can be done only by
reversible equations!}\,.

Realization of such a program just would give the answer to the
question. On this way, one can easy immediately see a mechanism of
generation of the spatial correlations and destroying the
 Sto{\ss}zahlansatz.

Making the first step, let us rewrite the second of BBGKYE in the
``pseudo-Liouville'' form \cite{pg}:
\begin{eqnarray}
\frac {\partial F_{2}}{\partial t}= \,a^2\!\!\int ({\bf v}_{12}\cd
\Omega)\, \delta({\bf r}_{12}-a\Omega)\, F_2\, d\Omega\,
-  \,\, \,\,\,\,\,\, \nonumber\\
- {\bf v}_{12}\cd \frac {\partial F_{2}}{\partial {\bf r}_{12}}-
\frac {{\bf v}_1\!+\!{\bf v}_2}{2}\cd \frac {\partial F_{2}}{\partial
{\bf R}}  +\nu\! \sum_{j\,=1}^{2} \widehat{\mathbf{I}}_{j,3}\,\,
F_{3} \,\,
 \, \label{f22}
\end{eqnarray}
Here, the coordinates $\,{\bf r}_j\,$ may enter the forbidden region
$\,|{\bf r}_{1}-{\bf r}_{2}|<a\,$, where $\,F_2\equiv 0\,$,\, and the
new (first on r.h.s.) term represents a force of repulsion of
particles at the border of this region\,\footnote{\, In place of\,
$\,(\nabla_{12}\Phi({\bf r}_{12})) \cdot (\partial F_2/\partial {\bf
p}_{1}-\partial F_2/\partial {\bf p}_{2})\,$\, in case of a smooth
interaction potential $\,\Phi(\rho )\,$.}\,. Besides, we separated
the relative displacement of particles, with $\,{\bf r}_{12}={\bf
r}_{1}-{\bf r}_{2}\,$, and motion of their center of mass,  $\,{\bf
R}=({\bf r}_{1}+{\bf r}_{2})/2\,$. The first of these two in turn can
be divided into norma; and tangential components:
\begin{eqnarray}
- {\bf v}_{12}\cd \frac {\partial F_{2}}{\partial {\bf r}_{12}}= -
({\bf v}_{12}\cd \Omega) \frac {\partial F_{2}}{\partial \rho_{12}}
- \left(\frac {{\bf v}_{12}}{\rho_{12}}\, \Lambda(\Omega)\,\frac
{\partial F_{2}}{\partial \Omega}\right) ,\, \label{nt}
\end{eqnarray}
where\, $\,\Omega ={\bf r}_{12}/|{\bf r}_{12}|\,$\,,\,
$\,\rho_{12}=|{\bf r}_{12}|\,$\, and\,
\begin{eqnarray}
\Lambda(\Omega)\,{\bf f} = {\bf f}- \Omega\,(\Omega \cd {\bf f}) = -
\left[  \Omega  \times  \left[\Omega\times {\bf f}\,\right] \right
]\,\nonumber
\end{eqnarray}

Next, consider, with the help of (\ref{f22}), the contact DF
$\,F_2({\bf r}_1={\bf r}_2+a\Omega) \equiv F_2^{(c)}(t,{\bf
R},\Omega,{\bf p}_1,{\bf p}_2)\,$\,, taking in mind that the first
term on r.h.s. of (\ref{f22}) and the first (normal) component of
(\ref{nt}) have quite similar, but oppositely signed, singularities
which compensate one another. Let us assume that this is exact
compensation, that is
\begin{equation}
a^2\!\!\int \! ({\bf v}_{12} \Omega^{\prime})\,
\delta(\rho_{12}\Omega-a\Omega^{\prime})\, F_2\, d\Omega^{\prime}
-({\bf v}_{12} \Omega) \frac {\partial F_{2}}{\partial \rho_{12}}
=0\,\label{ass}
\end{equation}
Then from (\ref{f22})-(\ref{ass}) the necessary autonomous evolution
equation for the pair contact DF does follow:
\begin{eqnarray}
\frac {\partial F_{2}^{(c)}}{\partial t}= - \frac {{\bf v}_1\!+\!{\bf
v}_2}{2}\cd \frac {\partial F_{2}^{(c)}}{\partial {\bf R}} -\,\,\,\,
\,\,\,\,
\,\,\,\, \,\,\,\, \,\,\,\, \,\,\,\, \label{f2c}\\
\,\,\,\, \,\,\,\, \,\,\, \,\, \,\,\, \,\, \,\,\,\,-\left(\frac {{\bf
v}_{12}}{a}\, \Lambda(\Omega)\,\frac {\partial F_{2}^{(c)}}{\partial
\Omega}\right)
+\nu\! \sum_{j\,=1}^{2} \widehat{\mathbf{I}}_{j,3}\,\, F_{3}^{(c)}
\,\,  \,, \nonumber
\end{eqnarray}
where $\,F_3^{(c)}\,$ is the mentioned contact DF for two ``bound
together'' pair collisions.

Notice that formally (\ref{ass}) is not assumption but identity. It
expresses continuity f probability distribution at breaks of phase
trajectories because of collisions, that is the same as the condition
(\ref{mr}) does express. In essence, this is analogue of equalities
(3)-(4) from \cite{i1} for a gas with smooth
interaction\,\footnote{\, Or, to be more concrete, analogue of the
equality \newline $\,(\nabla_{12}\Phi({\bf r}_{12}))  \cdot (\partial
F_2/\partial {\bf p}_{1} - \partial F_2/\partial {\bf p}_{2}) - ({\bf
v}_{12}\cdot \nabla_{12})\,F_2\,=\,0\,$ \newline which should be
satisfied (as identity or as ``ansatz'') in the space region occupied
by collision \cite{i1} (for instance, inside the ``collision
cylinder'' \cite{tmf,p3}), in order to equalize probabilities of
mutually corresponding\,{\it in\,}-\, and \,{\it out\,}-states.}.\,

From the equation (\ref{f2c}) it it clear that any spatial
inhomogeneity, which induces $\,\partial F_1^{(c)}/\partial {\bf r}_1
\neq 0\,$ and
 $\,\partial F_2^{(c)}/\partial {\bf R}   \neq 0\,$, automatically
excludes possibility of reduction of $\,F_2^{(c)}\,$ to  $\,F_1\,$
and thus BBGKYE to BE, absolutely independently on value of
$\,\mu\,$\,\,\footnote{\, In other words, the inhomogeneity works as
a source of pre-collision inter-particle correlations. And this is
not surprising:\, regardless of velocity of a given particle, in its
vicinity with linear size $\,\sim\lambda\,$ there is always\, $\,\sim
(\pi a^2/\lambda^2)*$ $*\lambda^3\nu =1\,$\,\, particles forming
an\,{\it in\,}-state together with the given one and thus
kinematically suitable for pre-collision correlation with it.}\,.

Notice also that, firstly, the equation (\ref{f2c}) is reversible and
besides invariant in respect to replacing $\,{\bf p}_{1},{\bf
p}_{2}\,$ by $\,{\bf p}_{1}^*,{\bf p}_{2}^*\,$, as it should be
according to (\ref{mr}) (thus, the function of (\ref{mr}) now is
extension of this symmetry property from equations to their
solutions).

Secondly, the equality (\ref{ass}) says, in particular,  that $\,
\partial F_{2}/\partial \rho_{12} =0\,$\, at $\,\rho_{12}=a+0\,$\,
and\, $\,({\bf v}_{12} \Omega)\neq 0\,$. This is natural analogue of
behavior of gas density nearby a flipping (in accordance with
(\ref{mr})) surface.

{\bf 6}.\, The contact DF $\,F_2^{(c)}\,$ serves as a measure of mean
(ensemble averaged) number density of pair collisions. It is clear
that it drifts with the center of mass velocity of colliding
particles. Similarly, DFs $\,F_s^{(c)}\,$ will drift with velocities
$\,({\bf v}_1 +...+{\bf v}_s)/s\,$. Therefore they are mutually
independent. All together they determine statistics of key
$\,s$-particle configurations which produce all other configurations
and eventually evolution of $\,F_1\,$\,\,\footnote{\, In more detail,
on such kind of DFs see \cite{i1,i2,p12}. To avoid misunderstandings,
it is useful to underline that the smallness of probabilities of
contacts of two or more particles in no way means smallness of
corresponding contact DFs since the latter (as well as all the DFs
under consideration) represent density of probability.}\,.

{\bf 7}.\, Formulation of equations for  $\,F_3^{(c)}\,$,
$\,F_4^{(c)}\,$, etc. we leave for the future. At present, it is more
important to point out existence of alternative, just discussed,
treatment of the hard sphere BBGKY hierarchy. This treatment, in
contrast with the conventional one, does not ignore the fundamental
reversibility property of the BBGKY equations but uses it as basis of
definite constructive approach to solutions of these equations. A
qualitatively similar approach to a gas of ``soft spheres'' was
tested in cite{i1,i2,p12}. The model of ``hard spheres'' is of
special interest because, expectedly, it can more easily achieve
formal quantitative rigor.

\end{document}